\documentclass[aps,pra,twocolumn,superscriptaddress,showpacs]{revtex4}
\usepackage{amssymb}
\usepackage{amsmath}
\usepackage[dvips]{graphicx}
\usepackage{subfigure}

\begin{document}

\title{Teleportation of entangled states without Bell-state measurement}
\author{Wesley B. Cardoso}
\affiliation{Instituto de F\'{\i}sica, Universidade Federal de Goi\'{a}s, 74.001-970, Goi\^{a}nia (GO), Brazil}
\author{A.T. Avelar}
\affiliation{Instituto de F\'{\i}sica, Universidade de Bras\'{\i}lia, 70.919-970, Bras%
\'{\i}lia (DF), Brazil}
\affiliation{Instituto de F\'{\i}sica, Universidade Federal de Goi\'{a}s, 74.001-970, Goi%
\^{a}nia (GO), Brazil}
\author{B. Baseia}
\affiliation{Instituto de F\'{\i}sica, Universidade Federal de Goi\'{a}s, 74.001-970, Goi%
\^{a}nia (GO), Brazil}
\author{N.G. de Almeida}
\email[Corresponding author: ]{nortongomes@ucg.br (NG de Almeida)}
\affiliation{N\'{u}cleo de Pesquisas em F\'{\i}sica, Universidade Cat\'{o}lica de Goi\'{a}%
s, 74.605-220, Goi\^{a}nia (GO), Brazil.}
\affiliation{Instituto de F\'{\i}sica, Universidade Federal de Goi\'{a}s, 74.001-970, Goi%
\^{a}nia (GO), Brazil}

\begin{abstract}
In a recent paper [Phys. Rev. A \textbf{70}, 025803 (2004)] we presented a
scheme to teleport an entanglement of zero- and one-photon states from a
bimodal cavity to another one, with $100\%$ success probability. Here,
inspired on recent results in the literature, we have modified our
previous proposal to teleport the same entangled state without using
Bell-state measurements. For comparison, the time spent, the fidelity, and
the success probability for this teleportation are considered.
\end{abstract}

\maketitle

Quantum teleportation, first suggested by Bennett \textit{et al} \cite%
{Bennett93}, is one of the cornerstones of the quantum computation \cite%
{Gottesman99} and quantum communication \cite{Cirac95,Brassard98}. The
crucial ingredient characterizing this phenomenon is the
transfer of information between non interacting systems, at the expense of a
quantum channel. It has received great attention since
its pioneer proposal, mainly after its experimental realizations, e.g.: by Bouwmeester \textit{et al}. \cite{Bouwmeester97}, Boschi \textit{et
al.} \cite{Boschi98}, Lombardi \textit{et al}. \cite{Lombardi02}, all them using pairs of entangled photons by the process of
parametric down-conversion; Furusawa \textit{et al.} \cite{Furusawa98},
using entangled squeezed states; D. Fattal \textit{et al.} \cite{Fattal04}, using a quantum dot single-photon source. Since then, various schemes have been
suggested to implement the teleportation process in different contexts, such
as trapped ions \cite{Solano01}, running wave fields \cite{Braunstein98,Villas99,Lee00,Lee01,Serra02}, and trapped wave fields inside
high-Q cavities \cite{Davidovich94,Moussa97,Pires04}.

To implement teleportation between two distinct separated points,
usually named as Alice and Bob, one should: i) prepare a state to be
teleported; ii) prepare the nonlocal quantum channel; iii) make a Bell
measurement (Alice); iv) communicate (to Bob), by classical channel, the
measured result. The main experimental challenge consists in the
so-called Bell state measurement \cite{Bennett93}, performed on the Bell operator basis for the particle (or field) whose state is to be teleported, plus its partner composing the quantum channel. The same procedure is employed when we are
concerned with teleportation of entangled states with major interest for
quantum information processing \cite{Moussa97,Loock00,Geisa}.

Various schemes for teleportion differing from the original protocol - in the sense that Bell-state measurement is not employed - have been proposed: in Ref.\cite{Vaidman}, Vaidman considered both a spin state and a system with continuous variable, and presented a \textquotedblleft cross measurement\textquotedblright\ method, thus obtaining a two-way teleportation. In Ref. \cite{napolitano}, de Almeida \textit{et al.} proposed a scheme to teleport a superposition of coherent states from a high$-Q$ cavity to another with $100\%$ fidelity. The method includes damping effects and creation of the state to be teleported. In Ref.\cite{Zheng04}, Zheng refers to the (approximated) teleportation of the superposition of zero- and one-photon states from a high$-Q$ cavity to another, with fidelity near $99\%$. Both the procedures in Refs.\cite{napolitano,Zheng04} considered a single mode of a trapped field interacting with a single two-level atom via the Jaynes-Cummings Hamiltonian. Soon after Ye and Guo \cite{Guo04} have treated the teleportation of an unknown atomic state in cavity QED; accordingly, the advantage of this scheme is that only virtual field excitations occur with the passage of atoms through the cavity, hence no transfer of quantum information happens between the atoms and the cavity when the cavity is initially assumed in the vacuum state. Also, as in Ref.\cite{napolitano}, the state teleportation without Bell state measurements occurs exactly, with $50\%$ of success probability.

Meanwhile \cite{Pires04}, a proposal for teleporting an entanglement of
zero- and one-photon states was presented with $100\%$ success probability and $100\%$ fidelity. This scheme requires a collection of two kinds of two-level atoms, a three-level atom in a ladder configuration driven by a classical field,
Ramsey zones plus selective atomic detectors. As usually,
Bell-measurements were employed. Here we will neglect the use of Bell measurements, as done in Refs.\cite{Vaidman,napolitano,Zheng04,Guo04}, to simplify the teleportation scheme of Ref.\cite{Pires04}. Our producere is inspired on the Ref.\cite{Zheng04} to implement an approximated conditional teleportation of an unknown entanglement of zero- and one-photon states from a bimodal high-Q cavity to another such one. This scheme employs two two-level (resonant, Rydberg) atoms, Ramsey zones and a selective atomic state detector. The success probability of this scheme coincides with that of the original protocol ($25\%$) if we restrict the measurements on the Bell basis to only one of the four Bell states. Differently from \cite{Pires04}, an additional atom is no longer used, which simplifies the teleportation of zero- and one-photon entanglement states. As assumed in Ref. \cite{Pires04,Zheng04}, the whole losses due to atomic spontaneous emission and dissipation in the cavities were neglected. Actually, since the decoherence time is of the same order of the lifetimes for the qubits in a high$-Q$ cavity and the (spontaneous) atomic decay, the experimental implementation of the present scheme should be made in a time interval lesser than the $10^{-2}s$, a typical time for both decoherence and damping of atomic and cavity qubits \cite{brune}.

Fig.1 displays the setup of experimental proposal: S$_{A}$ represents
\textquotedblleft Source of Atoms\textquotedblright , \textquotedblleft
Excitation\textquotedblright\ prepares a highly excited (Rydberg) atom,
\textquotedblleft C$_{1}$\textquotedblright\ (\textquotedblleft C$_{2}$%
\textquotedblright ) represents the first (second) bimodal cavity and $D_{e}$
$(D_{g})$ stands for atomic ionization detector for the state $|e\rangle $ $%
(|g\rangle )$. The two bimodal cavities sustain two nondegenerate orthogonally polarized modes \cite{Rauschenbeutel01}, \textit{A} and \textit{B} in C$_{1}$ (\textit{C} and \textit{D }in C$_{2}$). 
\begin{figure}[h]
\centering
\includegraphics[{height=2.5cm,width=8cm}]{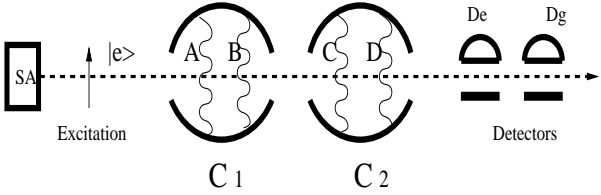}
\caption{Scheme of the experimental setup required for teleportation of an
entanglement of zero- and -one photon state without using Bell states.}
\label{cavidade}
\end{figure}

To perform the teleportation of zero- and one-photon entangled states
between two high-Q cavities we assume the second cavity $C_{2}$ initially
prepared in the state \cite{Pires04} 
\begin{equation}
\left\vert \psi \right\rangle _{C_{2}}=\alpha \left\vert 1\right\rangle
_{C}\left\vert 0\right\rangle _{D}+\beta \left\vert 0\right\rangle
_{C}\left\vert 1\right\rangle _{D},
\end{equation}%
where $\alpha $ and $\beta $ are unknown coefficients. First, we send a
two-level atom, in an excited state $|e\rangle _{1}$, through the initially
empty cavity $C_{1}$. The atom$-1$ interacts resonantly with mode $A$
according to the Jaynes-Cummings Hamiltonian \cite{Scully}. We note that
when interaction between the atom and one of the two modes takes place, the
other mode remains unaffected, as shown in Ref.\cite{Pires04}. We have, for
the \textit{on-resonant} interaction, 
\begin{equation}
H_{on}=\hbar g\left( \sigma^{-}a^{\dagger}+\sigma^{+}a\right) ,
\label{hon}
\end{equation}%
where $a^{\dagger}$ and $a$ stand for the creation and annihilation
operator for the cavity modes $A,B,C,D$; the $\sigma $'s represent the Pauli
operators; and $g$ is the atom-field coupling parameter. The state
describing the system in $C_{1}$ after the atom-field interaction evolves
from the initial state $|0\rangle _{A}|0\rangle _{B}|e\rangle _{1}$ to 
\begin{equation}
|\varphi (t)\rangle =\cos (gt)|0\rangle _{A}|0\rangle _{B}|e\rangle
_{1}-i\sin (gt)|1\rangle _{A}|0\rangle _{B}|g\rangle _{1},
\end{equation}%
and setting $gt=\pi /4$, we obtain 
\begin{equation}
|\varphi (\pi /4g)\rangle =\frac{1}{\sqrt{2}}\left( |0\rangle _{A}\left\vert
0\right\rangle _{B}\left\vert e\right\rangle _{1}-i\left\vert 1\right\rangle
_{A}\left\vert 0\right\rangle _{B}\left\vert g\right\rangle _{1}\right) .
\end{equation}%
Now, the state describing the system, including the cavity C$_{2}$, is 
\begin{eqnarray}
\left\vert \Psi \right\rangle &=&\frac{1}{\sqrt{2}}(\left\vert
0\right\rangle _{A}\left\vert 0\right\rangle _{B}\left\vert e\right\rangle
_{1}-i\left\vert 1\right\rangle _{A}\left\vert 0\right\rangle _{B}\left\vert
g\right\rangle _{1}) \nonumber\\
&& \otimes \left( \alpha \left\vert 1\right\rangle _{C}\left\vert
0\right\rangle _{D}+\beta \left\vert 0\right\rangle _{C}\left\vert
1\right\rangle _{D}\right) .
\end{eqnarray}

After the evolution due to the resonant interaction of atom$-1$ with mode $C$ of
cavity C$_{2}$ governed by the Hamiltonian (2), the state describing the whole system results in%
\begin{eqnarray}
\left\vert \Phi (t^{\prime })\right\rangle & =&\frac{\alpha }{\sqrt{2}}\cos
\left( \sqrt{2}gt^{\prime }\right) \left\vert 0\right\rangle _{A}\left\vert
0\right\rangle _{B}\left\vert 1\right\rangle _{C}\left\vert 0\right\rangle
_{D}\left\vert e\right\rangle _{1} \nonumber \\
& -&\frac{i\alpha }{\sqrt{2}}\sin \left( \sqrt{2}gt^{\prime }\right)
\left\vert 0\right\rangle _{A}\left\vert 0\right\rangle _{B}\left\vert
2\right\rangle _{C}\left\vert 0\right\rangle _{D}\left\vert g\right\rangle
_{1} \nonumber \\
& -&\frac{i\alpha }{\sqrt{2}}\cos \left( gt^{\prime }\right) \left\vert
1\right\rangle _{A}\left\vert 0\right\rangle _{B}\left\vert 1\right\rangle
_{C}\left\vert 0\right\rangle _{D}\left\vert g\right\rangle _{1} \nonumber \\
& -&\frac{\alpha }{\sqrt{2}}\sin \left( gt^{\prime }\right) \left\vert
1\right\rangle _{A}\left\vert 0\right\rangle _{B}\left\vert 0\right\rangle
_{C}\left\vert 0\right\rangle _{D}\left\vert e\right\rangle _{1} \nonumber \\
& +&\frac{\beta }{\sqrt{2}}\cos \left( gt^{\prime }\right) \left\vert
0\right\rangle _{A}\left\vert 0\right\rangle _{B}\left\vert 0\right\rangle
_{C}\left\vert 1\right\rangle _{D}\left\vert e\right\rangle _{1} \nonumber \\
& -&\frac{i\beta }{\sqrt{2}}\sin \left( gt^{\prime }\right) \left\vert
0\right\rangle _{A}\left\vert 0\right\rangle _{B}\left\vert 1\right\rangle
_{C}\left\vert 1\right\rangle _{D}\left\vert g\right\rangle _{1} \nonumber \\
& -&\frac{i\beta }{\sqrt{2}}\left\vert 1\right\rangle _{A}\left\vert
0\right\rangle _{B}\left\vert 0\right\rangle _{C}\left\vert 1\right\rangle
_{D}\left\vert g\right\rangle _{1}.
\end{eqnarray}%
When detecting the atom$-1$ in the state $\left\vert e\right\rangle _{1}$, $%
\ \left\vert \Phi (t^{\prime })\right\rangle $ collapses in the form%
\begin{eqnarray}
\left\vert \Phi (t^{\prime })\right\rangle _{col}& =&N_{1}\left[
\alpha \cos \left( \sqrt{2}gt^{\prime }\right) \left\vert 0\right\rangle
_{A}\left\vert 0\right\rangle _{B}\left\vert 1\right\rangle _{C}\left\vert
0\right\rangle _{D}\right. \nonumber \\
&-& \alpha \sin \left( gt^{\prime }\right) \left\vert 1\right\rangle
_{A}\left\vert 0\right\rangle _{B}\left\vert 0\right\rangle _{C}\left\vert
0\right\rangle _{D} \nonumber \\
&+& \left. \beta \cos \left( gt^{\prime }\right) \left\vert 0\right\rangle
_{A}\left\vert 0\right\rangle _{B}\left\vert 0\right\rangle _{C}\left\vert
1\right\rangle _{D}\right] .
\end{eqnarray}%
where $N_{1}$ is a normalization factor. 

Next the atom-2, previously prepared in the state $\left\vert e\right\rangle
_{2}$, interacts with the mode $B$ in C$_{1}$, leaving the entangled state
of the whole system in the form (with $gt=\pi /4$),%
\begin{eqnarray}
\left\vert \chi (\pi /4g)\right\rangle & = & \alpha [
cos(\sqrt{2}gt^{\prime})|0\rangle_{A}|0\rangle_{B} |1\rangle_{C}|0\rangle_{D}|e\rangle_{2} 
\nonumber \\
&&-\phantom{a}icos(\sqrt{2}gt^{\prime})|0\rangle_{A}|1\rangle_{B} |1\rangle_{C}|0\rangle_{D}|g\rangle_{2}
\nonumber \\
&&-\phantom{a}sin(gt^{\prime})|1\rangle_{A}|0\rangle_{B}|0\rangle_{C} |0\rangle_{D}|e\rangle_{2}
\nonumber \\
&&+ \phantom{a}\left. isin(gt^{\prime})|1\rangle_{A}|1\rangle_{B}|0\rangle_{C}|0\rangle_{D}|g\rangle_{2} \right]\nonumber \\
&+& \beta \left[ cos(gt^{\prime})|0\rangle _{A}|0\rangle_{B}|0\rangle_{C} |1\rangle_{D}|e\rangle_{2}\right. \nonumber \\
&&- \left. icos(gt^{\prime})|0\rangle_{A}|1\rangle_{B}|0\rangle_{C}|1\rangle_{D}|g\rangle_{2}\right] . \label{eq}
\end{eqnarray}
Then the atom$-2$ interacts with the mode $D$ in C$_{2}$ which leads the Eq.(\ref{eq}) to
\begin{eqnarray}
|\upsilon (t^{\prime })\rangle &=& \alpha\left[\cos(\sqrt{2}gt^{\prime}) \cos(gt^{\prime})|0\rangle_{A}|0\rangle_{B}|1\rangle_{C}|0\rangle_{D}|e\rangle_{2}
\right. \nonumber\\
&-& i\cos(\sqrt{2}gt^{\prime})\sin(gt^{\prime })
|0\rangle_{A}|0\rangle_{B}|1\rangle_{C}|1\rangle_{D}|g\rangle_{2} \nonumber \\
&-& i\cos(\sqrt{2}gt^{\prime})|0\rangle_{A}|1\rangle_{B}|1\rangle_{C} |0\rangle_{D}|g\rangle_{2} \nonumber \\ 
&+&i \sin^{2}(gt^{\prime})|1\rangle_{A}|0\rangle_{B}|0\rangle_{C} |1\rangle_{D}|g\rangle_{2}
\nonumber \\
&+&i \sin(gt^{\prime})|1\rangle_{A}|1\rangle_{B}|0\rangle_{C} |0\rangle_{D}|g\rangle_{2}
\nonumber \\
 &-& \left. \sin(gt^{\prime}) cos(gt^{\prime}) |1\rangle_{A}|0\rangle_{B}|0\rangle_{C}|0\rangle_{D}|e\rangle_{2}
\right] \nonumber \\
&+&\beta \left[ \cos(gt^{\prime})\cos(\sqrt{2}gt^{\prime})
|0\rangle_{A}|0\rangle_{B}|0\rangle_{C}|1\rangle_{D}|e\rangle_{2}\right. \nonumber \\
&-& i\cos(gt^{\prime})\sin(\sqrt{2}gt^{\prime}) \left\vert 0\right\rangle
_{A}\left\vert 0\right\rangle _{B}\left\vert 0\right\rangle _{C}\left\vert
2\right\rangle _{D}\left\vert g\right\rangle _{2} \nonumber \\
&-& i\cos^{2}(gt^{\prime })\left\vert 0\right\rangle _{A}\left\vert 1\right\rangle
_{B}\left\vert 0\right\rangle _{C}\left\vert 1\right\rangle _{D}\left\vert
g\right\rangle_{2} \nonumber \\
&-& \left. \cos(gt^{\prime})\sin(gt^{\prime})\left\vert 0\right\rangle _{A}\left\vert
1\right\rangle _{B}\left\vert 0\right\rangle _{C}\left\vert 0\right\rangle
_{D}\left\vert e\right\rangle _{2}\right] .
\end{eqnarray}%
Finally, detection of atom$-2$ in the state $\left\vert e\right\rangle _{2}$
projects the whole state as follows%
\begin{eqnarray}
\left\vert \psi \prime \right\rangle _{col} &=& N_{2}\left[
\alpha\cos(\sqrt{2}gt^{\prime})\cos(gt^{\prime })|0\rangle _{A}\left\vert
0\right\rangle _{B}\left\vert 1\right\rangle _{C}\left\vert 0\right\rangle_{D}\right. \nonumber \\
&+&\beta \cos(gt^{\prime})\cos(\sqrt{2}gt^{\prime }) \left\vert 0\right\rangle
_{A}\left\vert 0\right\rangle _{B}\left\vert 0\right\rangle _{C}\left\vert
1\right\rangle_{D} \nonumber \\
&-&\beta \cos(gt^{\prime})sin(gt^{\prime })\left\vert 0\right\rangle
_{A}\left\vert 1\right\rangle _{B}\left\vert 0\right\rangle _{C}\left\vert
0\right\rangle_{D} \nonumber \\
&-& \left. \alpha \sin(gt^{\prime })\cos(gt^{\prime})|1\rangle_{A}\left\vert
0\right\rangle _{B}\left\vert 0\right\rangle _{C}\left\vert 0\right\rangle
_{D}\right] .
\end{eqnarray}%
where $N_{2}$ is a normalization factor. 
Now, according to the protocol in
Ref.\cite{Zheng04}, we adjust $gt^{\prime }=7\pi /4$ implying $\cos \left( \sqrt{2}gt^{\prime }\right) \backsimeq 0.078 \backsimeq 0$, which leads the state in Eq.(10) to the form, 
\begin{equation}
\left\vert \psi \right\rangle _{C_{1}C_{2}}=\left( \alpha
\left\vert 1\right\rangle _{A}\left\vert 0\right\rangle _{B}+\beta 
\left\vert 0\right\rangle _{A}\left\vert 1\right\rangle _{B}\right) \otimes
\left\vert 0\right\rangle _{C}\left\vert 0\right\rangle _{D}
\end{equation}%
in which we recognize the teleportation of the initial state in the cavity C$%
_{2}$ (see Eq.1) to the cavity C$_{1}$, namely,
\begin{equation}
\left\vert \psi \right\rangle _{C_{1}}=\alpha \left\vert 1\right\rangle
_{A}\left\vert 0\right\rangle _{B}+\beta \left\vert 0\right\rangle
_{A}\left\vert 1\right\rangle _{B}.
\end{equation}%
This teleportation is attained with fidelity 
\begin{eqnarray}
F &=& \left\vert _{C_{1}C_{2}}\left\langle \psi \right. \left\vert \psi \prime \right\rangle
_{col}\right\vert^{2} \nonumber \\ 
&=& \frac{\sin^{2}(gt^\prime)}{cos^{2}(\sqrt{2}gt^{\prime})+sin^{2}(gt^{\prime})}\simeq 0.97. 
\end{eqnarray}
Note that the fidelity does not depend  on the arbitrary field coefficients $\alpha$ and $\beta$, as occurs in the Ref.\cite{Zheng04}. Here the direct measurements of the two atoms in the excited states, $\left\vert e\right\rangle _{1}$ and$\
\left\vert e\right\rangle _{2}$, substitute the Bell-state
measurements projecting a desired state in the cavity C$_{1}$ with fidelity $100\%$.

In summary, we have inspired by the scheme of Zheng \cite{Zheng04} to
teleport an arbitrary zero- and one-photon entangled states between
two bimodal high-Q cavities, using two two-level (Rydberg) atoms. Instead of obtaining exact teleportation, as found in the methods using the original protocol, here teleportation is attained approximately, with $25\%$ success probability (and $97\%$ fidelity), lesser than the $100\%$ (and $100\%$ ) found in \cite{Pires04}. However, in \cite{Pires04} the $100\%$ accuracy came from the use of two additional atoms to perform unitary operations by Bob, as required by the original protocol. By limiting the original protocol to only one of the four Bell states (and ignoring the atom preparing the state to be teleported), one reduces to $25\%$ the success probability of the scheme in \cite{Pires04} - while mantaining $100\%$ fidelity. Since the present procedure requires lesser number of atoms and, consequently, of selective atomic detectors, it becomes more economical than that of \cite{Pires04}: it reduces the time spent in the teleportation and decoherence effects caused by unavoidable interactions of the system with its environment, thus increasing its experimental feasibility. In additional, this time to complete the teleportation is of same order of that found in \cite{Zheng04}, both being lesser than the decoherence time.

We thank the CAPES (WBC) and the CNPq (ATA, BB, NGA), Brazilian Agencies,
for the partial supports of this work. NGA also thanks the Vice-Reitoria/UCG.

\end{document}